\documentclass[aps,reprint,twocolumn,amsmath,amssymb,showpacs]{revtex4-1}
\usepackage{graphicx,bm} \usepackage{color} 

\begin{document}

\title{Magnetizations and de Haas-van Alphen oscillations in massive Dirac fermions}

\author{F. R. Pratama} 
\email{pratama@flex.phys.tohoku.ac.jp}
\author{M.~Shoufie Ukhtary}
\author{Riichiro Saito}

\affiliation{Department of Physics, Tohoku University, Sendai
  980-8578, Japan}

\begin{abstract}
We theoretically study magnetic field, temperature, and energy band-gap dependences of magnetizations in the Dirac fermions. We use the zeta function regularization to obtain analytical expressions of thermodynamic potential, from which the magnetization of graphene for strong field/low temperature and weak field/high temperature limits are calculated. Further, we generalize the result by considering the effects of impurity on orbital susceptibility of graphene. In particular, we show that in the presence of impurity, the susceptibility follows a scaling law which can be approximated by the Faddeeva function. In the case of the massive Dirac fermions, we show that a large band-gap gives a robust magnetization with respect to temperature and impurity. In the doped Dirac fermion, we discuss the dependence of the band-gap on the period and amplitude of the de Haas-van Alphen effect. 


\end{abstract}

\date{\today}
\maketitle

\section{Introduction}


Historically, theoretical study on the magnetic properties of graphene~\cite{sepioni2010limits,goerbig2011electronic,fuseya2015transport} can be traced back to a paper by McClure~\cite{mcclure1956diamagnetism} in 1956. He showed that the diamagnetism in undoped graphene is largely contributed by coalescence of the massless Dirac electrons at the valence bands to the zeroth Landau levels (LLs) at the $K$ and $K^{\prime}$ valleys in the hexagonal Brillouin zone in the presence of an external magnetic field. At the zeroth LLs, the free energy increases with the increasing magnetic field, thus graphene shows orbital diamagnetism. Saito and Kamimura~\cite{saito1986orbital} showed that the orbital paramagnetism appears in graphene intercalation compounds. Furthermore, Raoux et al.~\cite{raoux2014from} demonstrated that the diamagnetism in a two-dimensional (2D) honeycomb lattice can be tuned into paramagnetism by introducing an additional hopping parameter, where the Berry phase is varying continuously between $\pi$ and $0$.

A method to derive analytical formula for orbital susceptibility of a massive Dirac system is developed by Koshino and Ando~\cite{koshino2010anomalous,koshino2011singular}. In their method, the Euler-Maclaurin expansion formula is applied to calculate the thermodynamic potential in the presence of the magnetic field. They showed that the pseudospin paramagnetism is responsible for a singular orbital susceptibility inside the band-gap region and disappears when the chemical potential enters the valence or conduction band~\cite{koshino2010anomalous,koshino2011singular,fukuyama2012dirac}. The expansion formula was first used by Landau~\cite{landau1930diamagnetismus} to demonstrate the orbital diamagnetism in metals. In the context of graphene-related materials, the Euler-Maclaurin expansion is employed to calculate orbital susceptibilities of  transition-metal dichalcogenides (TMDs)~\cite{cai2013magnetic} and the Weyl semimetals~\cite{koshino2016magnetic}. However, the magnetization as a function of magnetic field $B$ and temperature $T$ can not be obtained by using the Euler-Maclaurin expansion formula. It is because that the magnetization diverges due to an infinite number of the LLs formed in the valence bands  that areincluded in the calculation of the thermodynamic potential, unless a cut-off of the LLs is introduced~\cite{yan2017orbital}. The calculated magnetization $M$ as a function of $B$ has a form $M\propto C_1-C_2 B$ ($C_2>0$), where the constant $C_1$ becomes infinite with increasing the number of the LLs, while in the case of a conventional metal, only LLs in the conduction bands are considered. Moreover, the Euler-Maclaurin expansion method is valid only when the spacing of the LLs is much smaller than the thermal energy $k_B T$, where $k_B$ is the Boltzmann constant.~\cite{landau1930diamagnetismus}.

In the recent study by Li et al.~\cite{li2015field}, the magnetization of undoped graphene is  measured for a wide range of $B$ and temperature $T$. In the strong $B$/low $T$ limit, it is shown that the magnetization is proportional to the square-root of the magnetic field and diminishes linearly with increasing temperature ($M\propto-\sqrt{B}+\mathrm{constant}\times T$), while in the weak $B$/high $T$ limit, it is observed that the magnetization is proportional to $B$ and inversely proportional to $T$ ($M\propto-B/T$). The experimental data and numerical calculations are fitted into a Langevin function, from which the properties of magnetization for the two limiting cases can be deduced~\cite{li2015field}.     

To avoid the divergence in the magnetization, we derive analytical formula for thermodynamic potentials of the Dirac systems by using the zeta function regularization. In the context of quantum field theory, this method was used by Cangemi and Dunne~\cite{cangemi1996temperature} to calculate the energy of relativistic fermions in magnetic field. In graphene-related topics, the zeta function regularization was employed by Ghosal et al.~\cite{ghosal2007diamagnetism} to explain the anomalous orbital diamagnetism of graphene at $T=0~\mathrm{K}$ and by Slizovskiy and Betouras~\cite{slizovskiy2012nonlinear} to show the non-linear magnetization of graphene in a strong $B$ at high $T$. In this paper, we derive the analytical formula for magnetization of graphene for both strong $B$/low $T$ and weak $B$/high $T$ limits by the zeta function regularization, which reproduces the Langevin fitting to the experimental observation. Further, we discuss the effect of impurity on the orbital susceptibility of graphene. We show that in the presence of impurity, the orbital susceptibility as a function of temperature follows a scaling law which is approximately given by the so-called Faddeeva function. The effects of energy band-gap on the magnetization in undoped and doped Dirac systems are also discussed. In the undoped case, large band-gaps in TMDs yield relatively small but robust magnetizations with respect to temperature and impurity. In the doped case, we show that the opening of the band-gap is observed from the diminishing amplitude of the de Haas-van Alphen (dHvA) oscillation at $T=0~\mathrm{K}$. This phenomenon can not be obtained by the Euler-Maclaurin expansion method~\cite{mcclure1956diamagnetism}. 

The paper is organized as follows. In Sec. II, we present analytical methods for calculating the LLs, thermodynamic potential, and magnetization. In Sec. III, calculated results are discussed. In Sec. IV, conclusion is given.


\section{Calculation Methods}

\subsection{The Landau levels of massive Dirac fermions}

\begin{figure}[t]
\begin{center}
\includegraphics[width=8 cm, height=12.157 cm]{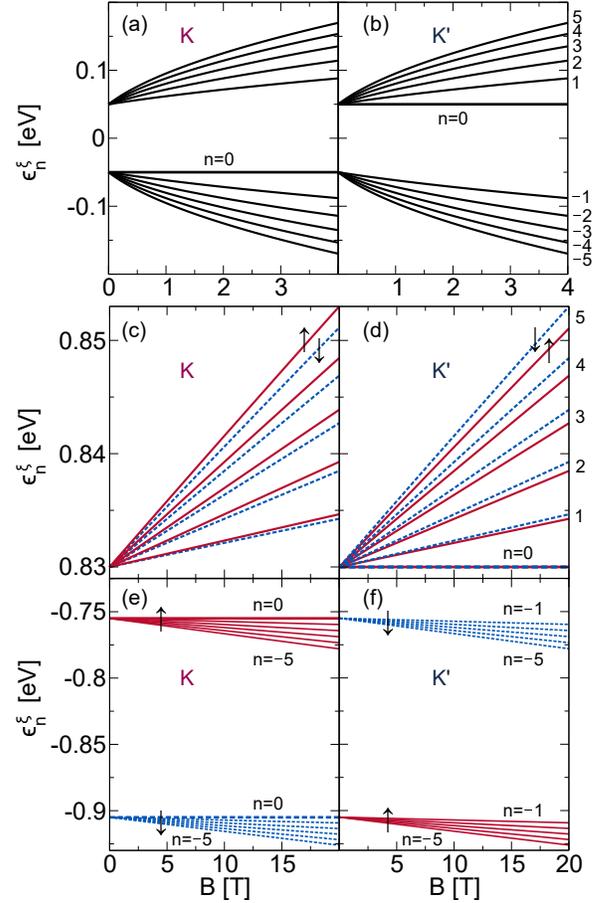}
\caption{ (Color online) The LLs ($n=-5$ to $n=5$) of the massive Dirac systems. The figures in the left (right) side correspond to the LLs at the $K$ ($K^{\prime}$) valley. (a), (b):  LLs of a gapped graphene ($\Delta=100~\mathrm{meV}$, $\lambda=0$, and $v_F=10^{6}~\mathrm{m/s}$) for $B=0-4~\mathrm{T}$. (c)-(f) The LLs of $\mathrm{MoS_2}$ ($\Delta=1.66~\mathrm{eV}$, $\lambda=75~\mathrm{meV}$ and $v_F=5.3\times 10^{5}~\mathrm{m/s}$) for $B=0-20~\mathrm{T}$. The LLs for the spin-up and spin-down electrons are shown with red solid and blue dashed lines, respectively.  }
\label{fig:Landaulvl}
\end{center}
\end{figure}

As a starting point, we consider a massive Dirac system with a band gap of $\Delta>0$. We employ a $2\times 2$ Hamiltonian matrix which is suitable to describe the energy spectra of a gapped graphene and transition-metal dichalcogenides (TMDs). The latter is enabled by including a non-zero spin-orbit coupling (SOC) constant $\lambda$ in the Hamiltonian~\cite{liu2013three}. The energy dispersions are approximated by a linear function of momentum $\textbf{p}=(p_x,p_y)$, and the Zeeman term is neglected because we only consider two cases: (1) at low temperature, the Zeeman splitting is much smaller than the Landau levels (LLs) separation and (2) at high temperature for TMDs. In the presence of an external magnetic field $\textbf{B}=B\hat{\textbf{z}}$, the momentum acquires an additional term by the Peierl substitution, i.e. $\textbf{p}\rightarrow\textbf{p}+e\textbf{A}$. The vector potential $\textbf{A}$ is related to $\textbf{B}$ by $\textbf{B}=\nabla\times\textbf{A}$. By choosing the Landau gauge $\textbf{A}=(0,Bx)$, the Hamiltonian is given by~\cite{goerbig2011electronic,fuseya2015transport,tse2011magneto}:
\begin{align}
\hat{H}_{\tau s}=
\begin{bmatrix}
\Delta/2  & v_{{F}}\{\tau p_x -i (p_y+eBx)\} \\
v_{{F}}\{\tau p_x +i (p_y+eBx)\} & -\Delta/2+\lambda\tau s
\end{bmatrix}.
\label{eq:peierl}
\end{align}
Here, $v_{F}$ is the Fermi velocity whose typical value for the Dirac fermions is $\sim 10^6~\mathrm{m/s}$, $\tau=+1~(-1)$ is the index for the $K$ ($K^{\prime}$) valley and $s=+1~(-1)$ is the index for spin-up (spin-down). 

To solve Eq.~(\ref{eq:peierl}), we define annihilation and creation operators $\hat{a}\equiv[\ell_B/(\sqrt{2}\hbar)][ip_x+(p_y+eBx)]$ and $\hat{a}^{\dagger}\equiv[\ell_B/(\sqrt{2}\hbar)][-ip_x+(p_y+eBx)]$, where $\ell_B=\sqrt{\hbar/(eB)}$ is the magnetic length. In term of the annihilation and creation operators, the Hamiltonian for a given $\tau$ and $s$ reduces to
\begin{align}
\hat{H}_{\xi}=
\begin{bmatrix}
\Delta/2  & -i\hbar\omega_c\hat{O}_\tau \\
i\hbar\omega_c\hat{O}_{\tau}^{\dagger} & -\Delta/2+\lambda\xi
\label{eq:landaumx}
\end{bmatrix},
\end{align}
where $\xi\equiv\tau s$, $\hat{O}_{+}\equiv\hat{a}$, $\hat{O}_{-}\equiv\hat{a}^{\dagger}$, and $\omega_c=\sqrt{2}v_{{F}}/\ell_B=\sqrt{2{v_{{F}}}^2 e B/\hbar}$ is the cyclotron frequency of the Dirac fermions. $\xi=+1(-1)$ represents the spin-up (spin-down) electron at the $K$ valley or spin-down (spin-up) electron at $K^{\prime}$ valley. The $n$-th LLs $\epsilon_n^{\xi}$ and wave function $|\Psi_{n}^{\xi}\rangle$ are given by the eigenvalues and eigenvectors of Eq.~(\ref{eq:landaumx}), respectively, as follows:
\begin{align}
\epsilon_n^{\xi} = \frac{\xi\lambda}{2} +\mathrm{sgn}_{\tau }(n)\sqrt{(\hbar\omega_c)^2|n|+\Big(\frac{\Delta_{\xi}}{2}\Big)^2}
\label{eq:landaulls}
\end{align}
and
\begin{align}
|\Psi_{n}^{\xi}\rangle=\frac{1}{\sqrt{2|\epsilon_n^{\xi}-\frac{\xi\lambda}{2}|}}
\begin{bmatrix}
-i{\sqrt{|\epsilon_n^{\xi}+\frac{\Delta}{2}-\lambda\xi|}} |\alpha_{n}^{\tau}\rangle\\
{\sqrt{|\epsilon_n^{\xi}-\frac{\Delta}{2}|}} |\beta_{n}^{\tau}\rangle
\end{bmatrix},
\label{eq:wavevector}
\end{align}
where we define $\Delta_{\xi}\equiv\Delta-\xi\lambda$ as a shorthand notation. $\mathrm{sgn}_{\tau }(n)$ is the sign function defined by $\mathrm{sgn}_{\tau }(n)=-1$ for $n<0$ and $\mathrm{sgn}_{\tau }(n)=+1$ for $n>0$. For $n=0$, a non-trivial wave function is satisfied by choosing $\mathrm{sgn}_{+}(0)=-1$ and $\mathrm{sgn}_{-}(0)=+1$. The eigenvectors $|\alpha_{n}^{\tau}\rangle$ and $|\beta_{n}^{\tau}\rangle$ are opposite for the $K$ and $K^{\prime}$ valleys, i.e. $|\alpha_{n}^{+}\rangle=|\beta_{n}^{-}\rangle\equiv||n|-1\rangle$ and $|\alpha_{n}^{-}\rangle=|\beta_{n}^{+}\rangle\equiv||n|\rangle$, where $||n|\rangle$ is a normalized eigenvector of the operators $\hat{a}^{\dagger}$ and $\hat{a}^{\dagger}$, such that $\hat{a}||n|\rangle=\sqrt{|n|}||n|-1\rangle$ and $\hat{a}^{\dagger}||n|\rangle=\sqrt{|n|+1}||n|+1\rangle$. The zeroth LLs at the $K$ and $K^{\prime}$ valleys exist at the valence and the conduction bands, respectively \cite{koshino2010anomalous,koshino2011singular}. The presence of only one zeroth LL in each valley is confirmed by first-principle calculations for hexagonal boron nitride (h-BN) and $\mathrm{MoS_2}$~\cite{lado2016landau}. For $\Delta<0$ and $\lambda=0$, a non-trivial wave function for $n=0$ is satisfied by $\mathrm{sgn}_{+}(0)=+1$ and $\mathrm{sgn}_{-}(0)=-1$, as in the case of topological silicene~\cite{tabert2013valley}. Nevertheless, in this study we only consider $\Delta>0$ without losing generality because it will be shown that the magnetization of the Dirac system depends only on the absolute value of $\Delta$, and not on the sign. It is noted that our convention of the $K$ and $K^{\prime}$ valleys is same as used in references~\cite{koshino2010anomalous,koshino2011singular,qu2017tunable} which is opposite of those references~\cite{cai2013magnetic,li2013unconventional}.

In Fig.~\ref{fig:Landaulvl}, we plot the LLs ($n=-5$ to $n=5$) of a gapped graphene [(a) and (b)] and $\mathrm{MoS_2}$ [(c)-(f)] at the $K$ and the $K^{\prime}$ valleys as a function of the magnetic field $B$. In (a) and (b), the LLs of the gapped graphene with $\Delta=100~\mathrm{meV}$, $\lambda=0$, and $v_F=10^{6}~\mathrm{m/s}$ show $\sqrt{B}$ dependences, because $\Delta/2$ is smaller than the cyclotron energy $\hbar\omega_c$ ($72.5~\mathrm{meV}$ for $B=4~\mathrm{T}$). In Fig.~\ref{fig:Landaulvl} (c)-(f), the LLs of $\mathrm{MoS_2}$ at the conduction bands [(c) and (d)] and the valence bands [(e) and (f)] are shown, where we adopt $\Delta=1.66~\mathrm{eV}$, $\lambda=75~\mathrm{meV}$ and $v_F=5.3\times 10^{5}~\mathrm{m/s}$~\cite{liu2013three,li2013unconventional,qu2017tunable,xiao2012coupled}. We can see that the SOC generates spin-splitting between the spin-up (red solid lines) and the spin-down (blue dashed lines) electrons except for the zeroth LLs at the $K^{\prime}$ valley [Fig.~\ref{fig:Landaulvl}(d)]. For the valence band, a spin-splitting $2\lambda=150~\mathrm{meV}$ occurs for the zeroth LLs at the $K$ valley [Fig.~\ref{fig:Landaulvl}(e)]. The LLs are linearly dependent for $B=0-20~\mathrm{T}$ because $\Delta/2$ is ten times larger than $\hbar\omega_c=0.086~\mathrm{eV}$ for $B=20~\mathrm{T}$.

\subsection{Thermodynamic potential and magnetization}

The thermodynamic potential per unit area at temperature $T$ in the presence of the magnetic field is given by
\begin{align}
\begin{split}
\Omega&=-\frac{1}{\beta}\frac{eB}{h}\sum_{\xi=\pm }\sum_{n=-\infty}^{\infty}\mathrm{ln}[1+e^{-\beta(\epsilon_n^{\xi}-\mu)}]\\
&\equiv\sum_{\xi=\pm }(\Omega_{-}+\Omega_{+}),
\end{split}
\label{eq:O}
\end{align}
where $\beta=1/(k_{{B}}T)$ and $\mu$ is the chemical potential. The pre-factor $eB/h$ represents the Landau degeneracy per unit area. For expository purposes, we define $\Omega_{-}$ and $\Omega_{+}$ as the potentials for the LLs at the valence and conduction bands, respectively. More specifically, $\Omega_{-}$ is the thermodynamic potential for the $n<0$ LLs and $n=0$ LL at the $K$ valley, and $\Omega_{+}$ is the thermodynamic potential for the $n>0$ LLs and $n=0$ LL at the $K^{\prime}$ valley.   The upper limit of $n$ depends on the magnitude of the LLs spacing compared with the thermal energy $k_B T$. When the LLs spacing is much larger than $k_B T$, the summation only includes the occupied LLs. When the LLs spacing is much smaller than $k_B T$, the summation can be taken for all LLs. In order to avoid divergence in Eq.~(\ref{eq:O}), the infinite summations of the LLs are expressed by the zeta function, from which a finite value from an infinite summation can be obtained through the method of analytical continuation. 

In the absence of SOC ($\lambda=0$), we drop the summation by the index $\xi$ because all the LLs are degenerate for the spin-up and spin-down electrons, and the thermodynamic potential is multiplied by $g_s=2$ to account the spin degeneracy. After obtaining an analytical expression of $\Omega$, the magnetization is calculated by
\begin{align}
M=-\frac{\partial\Omega}{\partial B}.
\label{eq:magnetization}
\end{align}

In the presence of impurity, the magnetization for a given scattering rate $\gamma$ is calculated by convolution of $M$ in Eq.~(\ref{eq:magnetization}) with a Lorentzian profile as follows \cite{sharapov2004magnetic,koshino2007diamagnetism,nakamura2007orbital,nakamura2008electric,tabert2015magnetic}:\begin{align}
M(\mu,\gamma)=\frac{\gamma}{\pi}\int_{-\infty}^{\infty}d\varepsilon M(\varepsilon) \frac{1}{(\varepsilon-\mu)^2+\gamma^2}.
\label{eq:impurity}
\end{align}
The parameter $\gamma$ is related to the self-energy due impurity scattering, and $\gamma$ is inversely proportional to the relaxation time of the quasiparticle. For simplicity, we assume that $\gamma$ is independent of $B$ and $T$, and therefore the susceptibility as a function of temperature is given by $\chi(\mu,\gamma,T)=[\partial M(\mu,\gamma,T)/\partial B]_{B=0}$.


\subsubsection{Thermodynamic potential for $\hbar\omega_c\gg k_B T,~\Delta_{\xi}\gg k_B T$}\label{2B1}

First, let us derive the thermodynamic potential for $\hbar\omega_c\gg k_B T$. Since we consider electron-doped system, we get $\mu\geqslant \Delta/2$. The logarithmic function in Eq.~(\ref{eq:O}) is approximated by $\mathrm{ln}[1+\mathrm{exp}\{-\beta(\epsilon_n^{\xi}-\mu)\}]\approx-\beta(\epsilon_n^{\xi}-\mu)$ which is valid in the case of $-\beta(\epsilon_n^{\xi}-\mu)\gg k_B T$ or $T\rightarrow 0~\mathrm{K}$. The thermodynamic potential for the occupied LLS with $n\leqslant 0$, $\Omega_{-}$ is expressed by (see Appendix \ref{A1} for derivation),
\begin{align}
\Omega_{-}=-2\frac{eB}{h}\sum_{\xi=\pm}\Bigg[\hbar\omega_c\zeta\Big(-\frac{1}{2},{\Gamma_{\xi}}^2\Big)-\frac{\Delta_\xi}{4} \Bigg],
\label{eq:OvlowT}
\end{align}
where we define $\Gamma_{\xi}\equiv\Delta_{\xi}/(2\hbar\omega_c)$, and the infinite summation of the LLs is given by the Hurwitz zeta function which is defined by~(see for example reference \cite{olver2010nist}),
\begin{align}
\zeta(p,q)\equiv\sum_{k=0}^{\infty}\frac{1}{(k+q)^{p}}.
\label{eq:zeta}
\end{align}
It is noted that the chemical potential $\mu$ does not appear in the expression of $\Omega_{-}$ for the electron-doped system. Eq.~(\ref{eq:OvlowT}) gives the intrinsic diamagnetism of TMDs at $T=0~\mathrm{K}$. A similar result was derived by Sharapov et al.~\cite{sharapov2004magnetic} for a gapped graphene, which is obtained by introducing an ultraviolet cut-off in the calculation of the thermodynamic potential. By using the asymptotic form of $\zeta(p,q)$ for $q\rightarrow\infty$ and $p\neq 1$,
\begin{align}
\zeta(p,q)\sim \frac{q^{1-p}}{p-1}+\frac{q^{-p}}{2}+\frac{1}{12}pq^{-(1+p)},
\label{eq:asymptotic}
\end{align}
the $\Omega_{-}$ of a heavy Dirac fermion ($\Delta_{\xi}/2\gg\hbar\omega_c$) is given by
\begin{align}
\Omega_{-}\approx\sum_{\xi=\pm}\Bigg(\frac{1}{24\pi}\frac{{\Delta_{\xi}}^3}{(\hbar v_F)^2}+\frac{eB}{h}\frac{(\hbar\omega_c)^2}{6\Delta_{\xi}}\Bigg).
\label{eq:Ovheavy}
\end{align} 

In the calculation of $\Omega_{+}^{(e)}$, we define an integer $\nu_{\xi}$ as the index of the highest occupied LLs in the conduction bands since the summations of the LLs in the thermodynamic potential are carried out up to the $\nu_\xi$-th levels, as follows: 
\begin{align}
\nu_{\xi}\equiv\Bigg\lfloor\frac{(\mu-\xi\lambda/2)^2-({\Delta_{\xi}}/2)^2}{(\hbar\omega_c)^2}\Bigg\rfloor\equiv\lfloor\tilde{\nu_{\xi}}\rfloor,
\label{eq:nuxi}
\end{align}
where the floor function $\lfloor x \rfloor$ is defined by the greatest integer smaller than or equal to $x$. It is noted that when $\lambda=0$, we drop subscript $\xi$ in Eq.~(\ref{eq:nuxi}). When we introduce the step function $\Theta(\mu-\Delta/2)$ as a threshold, we confirm that $\mu\geqslant\Delta/2$ is relevant for electron-doped case $\Omega_{+}^{(e)}$, as follows (see Appendix \ref{A2} for derivation):
\begin{align}
\begin{split}
\Omega_{+}^{(e)}
&=-2 \frac{eB}{h}\sum_{\xi=\pm }\Bigg[\frac{\Delta}{4}+\mu\Big(\nu_{\xi}+\frac{1}{2}\Big)-\frac{\xi\lambda}{2}(\nu_{\xi}+1)  \\
&~~~+\hbar\omega_c\Bigg\{\zeta\Big(-\frac{1}{2},{\Gamma_{\xi}}^2+\nu_{\xi}+1\Big)- \zeta\Big(-\frac{1}{2},{\Gamma_{\xi}}^2\Big)\Bigg\}\Bigg]\\
&~~~\times\Theta(\mu-\Delta/2).
\end{split}
\label{eq:OclowT}
\end{align}
In Eq.~(\ref{eq:OclowT}), we need the fact that the finite summation of the LLs  is expressed by a subtraction of two zeta functions as follows~\cite{olver2010nist}: 
\begin{align}
\sum_{n=0}^{N}(n+q)^{-p}=\zeta(p,q)-\zeta(p,q+N+1).
\label{eq:difference}
\end{align}
In Appendix B, we show that the numerical calculation of the left-hand side of Eq.~(\ref{eq:difference}) reproduces the analytical expression on the right-hand side. Therefore, for the electron-doped case, the total thermodynamic potential is given by the summation $\Omega_{-}+\Omega_{+}^{(e)}$ [Eqs. (\ref{eq:OvlowT}) and (\ref{eq:OclowT})].

In the case of hole-doped system, on the other hand $\nu_{\xi}$ is now defined as the highest occupied LLs in the valence bands. By using the same procedure as $\Omega_{+}^{(e)}$, $\Omega_{-}^{(h)}$ is given as follows:
\begin{align}
\begin{split}
\Omega_{-}^{(h)}&=\frac{2eB}{h}\sum_{\xi=\pm}\Bigg[\frac{\Delta}{4}+|\mu|\Big(\nu_{\xi}+\frac{1}{2}\Big) +\frac{\xi\lambda}{2}\nu_{\xi}  \\
 &~~~+\hbar\omega_c\Bigg\{\zeta\Big(-\frac{1}{2},{\Gamma_{\xi}}^2+\nu_{\xi}+1\Big)-\zeta\Big(-\frac{1}{2},{\Gamma_{\xi}}^2\Big)\Bigg\}\Bigg]\\
 &~~~\times\Theta(|\mu|-\Delta_{\xi}/2).
\end{split}
\label{eq:OhlowT}
\end{align}
The total thermodynamic potential is given by the subtraction $\Omega_{-}-\Omega_{-}^{(h)}$  [Eqs.(\ref{eq:OvlowT}) and (\ref{eq:OhlowT})] because $\Omega_{-}^{(h)}$ represents the "potential" of the unoccupied LLs. It is noted that the electron- and hole-doped Dirac systems give identical thermodynamic potentials in the case of $\lambda=0$, because of the electron-hole symmetry.

\subsubsection{Thermodynamic potential for $\hbar\omega_c\sim\Delta\ll k_B T$}\label{2B2}
Now, let us derive the thermodynamic potential of a gapped graphene ($\hbar\omega_c\sim\Delta$, $\lambda=0$) at high temperature. By considering $\mu>\Delta/2$, the total thermodynamic potential is given by $\Omega=\Omega_{-}+\Omega_{+}^{(e)}$. Here, we express $\Omega_{+}^{(e)}=\Omega_{+}-\Omega_{+}^{\prime}$, where $\Omega_{+}$ and $\Omega_{+}^{\prime}$ are the thermodynamic potentials from the entire LLs and from LLs $n=\nu$ to $n=\infty$ at the conduction bands, respectively. In Appendix \ref{A3}, we show that $\Omega_{+}^{\prime}$ is negligible for $\mu\ll k_B T$. We expand the logarithmic and exponential terms in the first line of Eq.~(\ref{eq:O}) to calculate $\Omega_{\mp}$ which is valid for $[-\epsilon_{\mp n}^{\xi}+\mu]\ll k_B T$. The thermodynamic potential $\Omega=\Omega_{-}+\Omega_{-}$ is given by a power series of $(\hbar\omega_c)^2$ as follows:
\begin{align}
\begin{split}
\Omega&=\frac{4g_s}{\beta}\frac{eB}{h}\sum_{\ell=0}^{\infty}{\mathrm{Li}_{1-2\ell}(-e^{\beta \mu })}\frac{(\beta \hbar \omega_c)^{2\ell}}{(2\ell)!}\Bigg[\zeta(-\ell,\Gamma^2)-\frac{\Gamma^{2\ell}}{2}\Bigg]\\
&\equiv\sum_{\ell=0}^{\infty}\Omega_{\ell},
\end{split}
\label{eq:OhighT2}
\end{align}
where $\Gamma=\Delta/(2\hbar\omega_c)$ for clarity. It is noted that the dependence of $\Omega$ on $\mu$ is given by the polylogarithm function $\mathrm{Li}_s(z)\equiv\sum_{k=1}^{\infty}z^{k}/k^{s}$, which converges for $|x|>1$ through the analytical continuation~\cite{olver2010nist} (see Appendix \ref{A3} for the derivation of Eq.~(\ref{eq:OhighT2})). By using the relation $\zeta(-\ell,x)=-\mathcal{B}_{\ell+1}(x)/(\ell+1)$~\cite{olver2010nist}, where $\mathcal{B}_{\ell}(x)$ is the Bernoulli polynomial, $\mathcal{B}_{1}(x)=x-1/2$ and $\mathrm{Li}_{1}(z)=-\mathrm{ln}(1-z)$~\cite{olver2010nist}, $\Omega_{0}$ is given by
\begin{align}
\Omega_{0}=k_B T\frac{g_s}{4\pi}\frac{\Delta^2}{(\hbar {v_F})^2}\mathrm{ln}[1+e^{\beta\mu}].
\label{eq:O0}
\end{align}
$\Omega_{0}$ is proportional to the square of band-gap $\Delta^2$ as well as temperature $T$ and does not depend on the magnetic field. Thus the $\Omega_0$ can be interpreted as the amount of thermal energy required to excite electrons from the valence bands across the band-gap.  $\Omega_{\ell}$ with $\ell\neq 0$ gives a linear response of the magnetization as a function of $B$, as shall be discussed in the next section.

\section{Results}

\subsection{Magnetization of graphene ($\Delta=0,~\lambda=0$)}

\begin{figure}[t]
\begin{center}
\includegraphics[width=7 cm, height=13.721 cm]{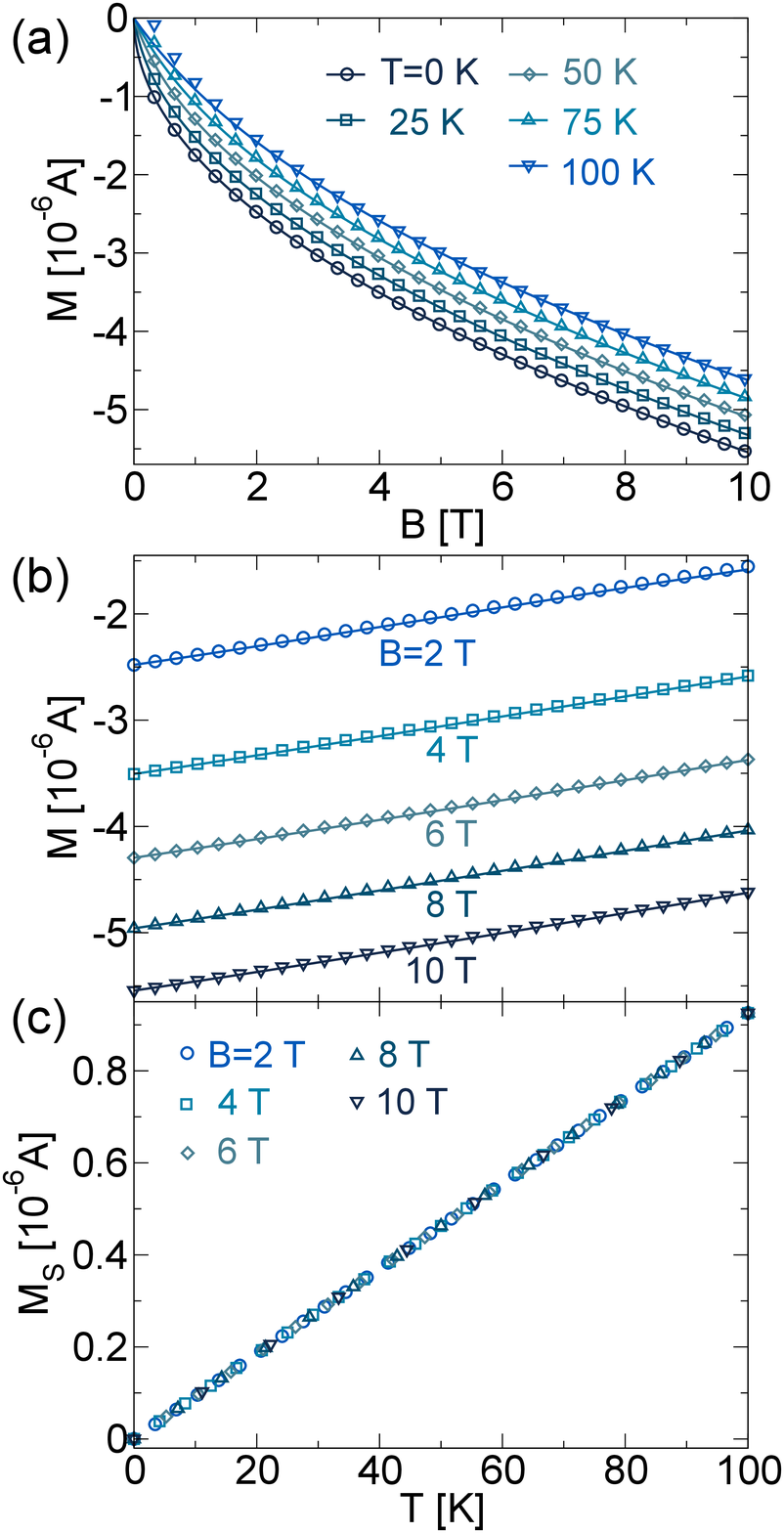}
\caption{ (Color online)  Magnetization of graphene at $\hbar\omega_c\gg k_B T$ limit as a function of magnetic field and temperature. (a) $M$ as a function of $B=0-10~\mathrm{T}$ at fixed temperatures, (b) $M$ and (c) $M_S$ as a function of $T$ ($T=0-100~\mathrm{K}$) for several values of $B$. The calculations from the analytical formula [Eq.~(\ref{eq:GrM})] and the Langevin function [Eq.~(\ref{eq:langevin})] are depicted by symbols and lines, respectively. }
\label{fig:GrlowT}
\end{center}
\end{figure}

\begin{figure}[t]
\begin{center}
\includegraphics[width=7 cm, height=12.998 cm]{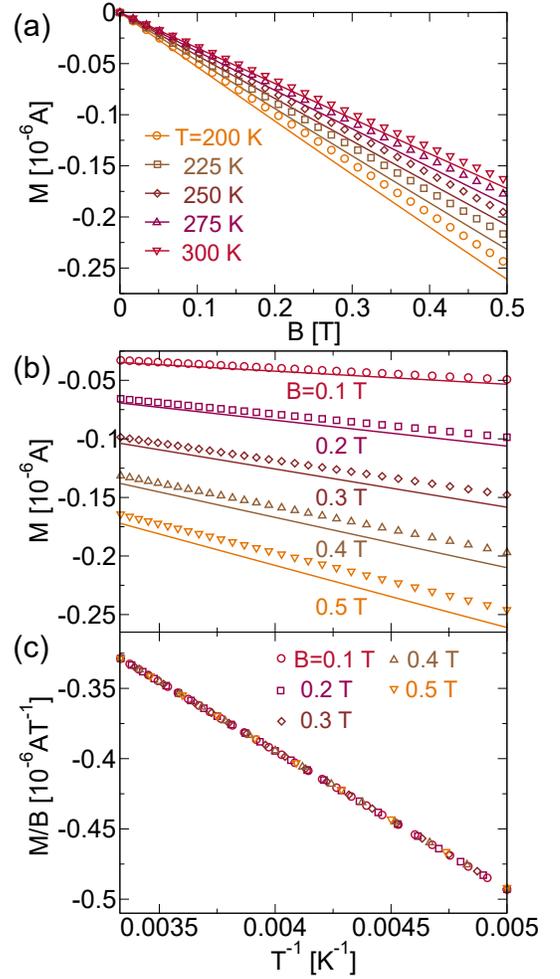}
\caption{ (Color online) Magnetization of graphene at $\hbar\omega_c\ll k_B T$ limit as a function of magnetic field and temperature. (a) $M$ as a function of  $B=0-0.5~\mathrm{T}$ at fixed temperatures, (b) $M$ and (c) $M/B$ as a function $1/T$ ($T=200-300~\mathrm{K}$) for several values of $B$. The calculations from the analytical formula [Eq.~(\ref{eq:GrM})] and the Langevin function [Eq.~(\ref{eq:langevin})] are depicted by symbols and lines, respectively.  }
\label{fig:GrhighT}
\end{center}
\end{figure}
In the case of graphene, the LL is given by $\epsilon_{n}^{\xi}=\mathrm{sgn}_{\tau}(n)=\hbar\omega_c\sqrt{n}$ and the zeroth LLs $\epsilon_{0}^{\xi}=0$ are shared between the valence and conduction bands at the $K$ and $K^{\prime}$ valleys, respectively. By separating the zeroth LLs from the $n\neq 0$ LLs, the thermodynamic potentials $\Omega_{-}$ and $\Omega_{+}$ for $\hbar\omega_c\gg k_B T$ are given by
\begin{align}
\begin{split}
\Omega_{-}&=-\frac{g_s}{\beta}\frac{eB}{h}\Bigg[\mathrm{ln}\{1+e^{\beta\mu}\}+2\sum_{n=1}^{\infty}\mathrm{ln}\{1+e^{\beta(\hbar\omega\sqrt{n}+\mu)}\}\\
&=-2g_s\frac{eB}{h}\Bigg[\frac{1}{2\beta}\mathrm{ln}\{1+e^{\beta\mu}\}+\hbar\omega_c\zeta\Big(-\frac{1}{2},1\Big)-\frac{\mu}{2}\Bigg],
\end{split}
\label{eq:GrOvlowT}
\end{align}
and
\begin{align}
\begin{split}
\Omega_{+}&=-\frac{g_s}{\beta}\frac{eB}{h}\Bigg[\mathrm{ln}\{1+e^{\beta\mu}\}+2\sum_{n=1}^{\nu}\mathrm{ln}\{1+e^{-\beta(\hbar\omega\sqrt{n}+\mu)}\}\Bigg]\\
&=-2g_s\frac{eB}{h}\Bigg[\frac{1}{2\beta}\mathrm{ln}\{1+e^{\beta\mu}\}+\mu\nu\\
&~~~~+\hbar\omega_c\Bigg\{\zeta\Big(-\frac{1}{2},1+\nu\Big)-\zeta\Big(-\frac{1}{2},1\Big)\Bigg\}\Bigg],
\end{split}
\label{eq:GrOclowT}
\end{align}
respectively. We can confirm that by putting $\Delta=\lambda=0$, and $T\rightarrow 0~\mathrm{K}$, Eqs.~(\ref{eq:OvlowT}) and (\ref{eq:OclowT}) reduce to Eqs.~(\ref{eq:GrOvlowT}) and (\ref{eq:GrOclowT}), respectively. In the undoped graphene ($\mu=0$), the total thermodynamic potential is therefore given by
\begin{align}
\begin{split}
\Omega(\mu=0)&=-2g_s\frac{eB}{h}\Bigg[\hbar\omega_c\zeta\Bigg(-\frac{1}{2},1\Bigg)+\frac{1}{\beta}\mathrm{ln}(2)\Bigg]\\
&\equiv\Omega_B+\Omega_S. 
\label{eq:GrOlowT}
\end{split}
\end{align}
Since the thermodynamic potential can be equivalently expressed by $\Omega=E-TS$, where $E$ and $S$ are, respectively, the internal energy and entropy, we define $\Omega_B$ and $\Omega_S$ in Eq.~(\ref{eq:GrOlowT}) to denote the thermodynamic potential associated with the $n<0$ LLs and the entropy $S$ of the zeroth LLs, respectively.

In the case of $\hbar\omega_c\ll k_B T$, the magnetization is given by $\Omega$ in Eq.~(\ref{eq:OhighT2}) by putting $\Delta=0$  as follows:
\begin{align}
\Omega=\frac{-4g_s}{\beta}\frac{eB}{h}\sum_{\ell=1}^{\infty}\mathrm{Li}_{1-2\ell}(-e^{\beta\mu})\frac{(\beta\hbar\omega_c)^{2\ell}}{(2\ell )!}\frac{\mathcal{B}_{\ell+1}}{\ell+1}.
\label{eq:GrOhighT}
\end{align}
Since $\mathrm{Li}_{-1}(z)=z/(1-z)^2$, we reproduce the formula for susceptibility of graphene derived by McClure~\cite{mcclure1956diamagnetism}
\begin{align}
\chi=-\frac{e^2{v_F}^2}{6\pi k_B T}\mathrm{sech}^2\Big(\frac{\mu}{2k_B T}\Big ).
\label{eq:mcclure}
\end{align}
It is noted that Eq.~(\ref{eq:mcclure}) is valid for any temperature $T>0~\mathrm{K}$ because we take $B=0$ to calculate $\chi$, thus the condition $\hbar\omega_c\ll k_B T$ is always satisfied.

From Eqs.~(\ref{eq:GrOlowT}) and (\ref{eq:GrOhighT}), the magnetization of undoped graphene ($\mu=0$) is given by 
\begin{align}
M=
\begin{cases}
-\displaystyle\frac{0.882}{\pi}\displaystyle\frac{e^{3/2}v_F}{\hbar^{1/2}}\sqrt{B}+\displaystyle\frac{2\mathrm{ln}(2)}{\pi}\displaystyle\frac{e}{\hbar}k_B T,~(\hbar\omega_c\gg k_BT),\\
~\\
-{0.167}\displaystyle\frac{e^2 {v_F}^2}{\pi}\displaystyle\frac{B}{k_B T}+\mathcal{O}(B^3),~(\hbar\omega_c\ll k_BT).
\end{cases}
\label{eq:GrM}
\end{align}
It is noted that in Eq.~(\ref{eq:GrM}) for $\hbar\omega_c\ll k_BT$, only odd powers of $B$ survive because the Bernoulli number of $\mathcal{B}_{\ell+1}$ is zero for even $\ell> 0$.

The analytical expressions given by Eq.~(\ref{eq:GrM}) can be directly compared with the work of Li et al.~\cite{li2015field}. In their study, numerical calculation and experimental measurement of the magnetization for undoped graphene as a function of $B$ and $T$ are fitted into a Langevin function $L(x)=\mathrm{coth}(x)-1/x$ as follows:
\begin{align}
M=-\frac{0.882}{\pi}\frac{e^{3/2}v_F\sqrt{B}}{\hbar^{1/2}}L\Bigg(\frac{\sqrt{\hbar v_{F}^2 e B \alpha(T)}}{\sqrt{2}k_B T}\Bigg).
\label{eq:langevin}
\end{align}
The temperature dependence of $M$ is approximated by the function $\alpha(T)\equiv C/(C+\sqrt{T})$, where $C=45~\mathrm{K^{1/2}}$. Since $L(x)\sim x/3$ as $x\rightarrow 0$ and saturated to $1$ as $x\rightarrow\infty$, the magnetization is given by~\cite{li2015field}:
\begin{align}
M\approx
\begin{cases}
-\displaystyle\frac{0.882}{\pi}\displaystyle\frac{e^{3/2}v_F}{\hbar^{1/2}}\sqrt{B}+\displaystyle\frac{0.882\sqrt{2}}{\pi}\displaystyle\frac{e}{\hbar}k_B T,~(\hbar\omega_c\gg k_BT),\\
~\\
-\displaystyle\frac{0.882}{3\sqrt{2}}\displaystyle\frac{e^2 {v_F}^2}{\pi}\displaystyle\frac{B}{k_B T},~(\hbar\omega_c\ll k_BT).
\end{cases}
\label{eq:langevin2}
\end{align}
By comparing Eq.~(\ref{eq:GrM}) with Eq.~(\ref{eq:langevin2}), the analytical formula reproduces experimentally observed $B$ and $T$ dependences of the magnetization of graphene both for $\hbar\omega_c\gg k_BT$ and $\hbar\omega_c\ll k_BT$. It suggest that the zeta regularization works reasonably.

In Fig.~\ref{fig:GrlowT}(a), we plot $M(B,T)$ for $\hbar\omega_c\gg k_BT$ as a function of $B$ for several values of $T$ by the analytical expression Eq.~(\ref{eq:GrM}) (symbols) and the Langevin function Eq.~(\ref{eq:langevin}) (lines). We can see that Eq.~(\ref{eq:GrM}) works well at temperature as high as $T=100~\mathrm{K}$ for $B\geqslant 1~\mathrm{T}$, but for $B<1~\mathrm{T}$ Eq.~(\ref{eq:GrM}) overestimates the temperature dependence of the magnetization because the condition $\hbar\omega_c\gg k_B T$ does not satisfy for $B<1~\mathrm{T}$. In Fig.~\ref{fig:GrlowT}(b) we show $M$ as a function of $T$ for several values of $B$. The linear dependence of $M$ on $T$ originates from the entropy $S$ of electrons which coalesce to the zeroth LLs [see Eq.~(\ref{eq:GrOlowT})]. In Fig.~\ref{fig:GrlowT}(c), we plot $M_S\equiv M+C\sqrt{B}$, with $C\equiv 0.882 e^{3/2} v_F / (\pi \hbar^{1/2})$ [see Eq.~(\ref{eq:GrM})] as a function of $T$ for several values of $B$, which is a 'paramagnetic' contribution from the entropy of the zeroth LLs. Here, the functions $M_S$ are aligned into a straight line, which shows that $M_S$ is independent of $B$.

In Fig. \ref{fig:GrhighT}(a), we plot $M(B,T)$ for $\hbar\omega_c\ll k_BT$ as a function of $B$ for several values of $T$ by the analytical expression Eq.~(\ref{eq:GrM}) (symbols) and the Langevin function Eq.~(\ref{eq:langevin}) (lines), where the linear $B$ dependences of $M$ for $B\leq 0.5~\mathrm{T}$ are observed at temperature as low as $T=200~\mathrm{K}$, especially for weak $B\sim 0.1~\mathrm{T}$. For stronger $B$, Eqs.~(\ref{eq:GrM}) and (\ref{eq:langevin}) begin to show some discrepancies. In Fig. \ref{fig:GrhighT}(b), $M$ is plotted as a function of $1/T$ for several values of $B$. In Fig. \ref{fig:GrhighT}(c), the function $M/B$ is plotted as a function of $1/T$ and is aligned into a straight line which illustrates $M\propto -B/T$ dependence.  From Eq.~(\ref{eq:OhighT2}), it is inferred that the linear response of $M$ with increasing $B$ is originated from the contribution of the entire LLs at the valence and conduction bands.

\subsection{Magnetization for massive Dirac fermions $T=0~\mathrm{K},~\lambda=0$}

\begin{figure}[t]
\begin{center}
\includegraphics[width=7 cm, height=4.955 cm]{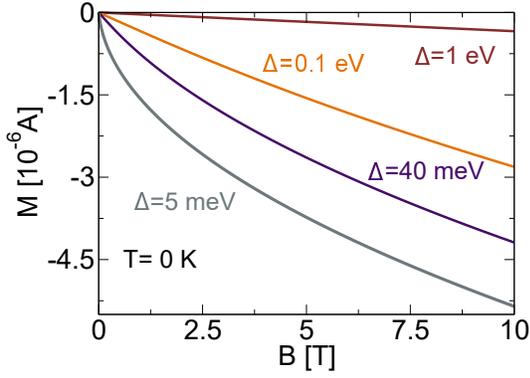}
\caption{ (Color online) Magnetization of massive Dirac fermions ($\Delta=5~\mathrm{meV},~40~\mathrm{meV},~0.1~\mathrm{eV}$ and $1~\mathrm{eV}$) as a function $B=0-10~\mathrm{T}$ at $T=0~\mathrm{K}$. }
\label{fig:DiracT0}
\end{center}
\end{figure}

\begin{figure}[t]
\begin{center}
\includegraphics[width=7 cm, height=10.42 cm]{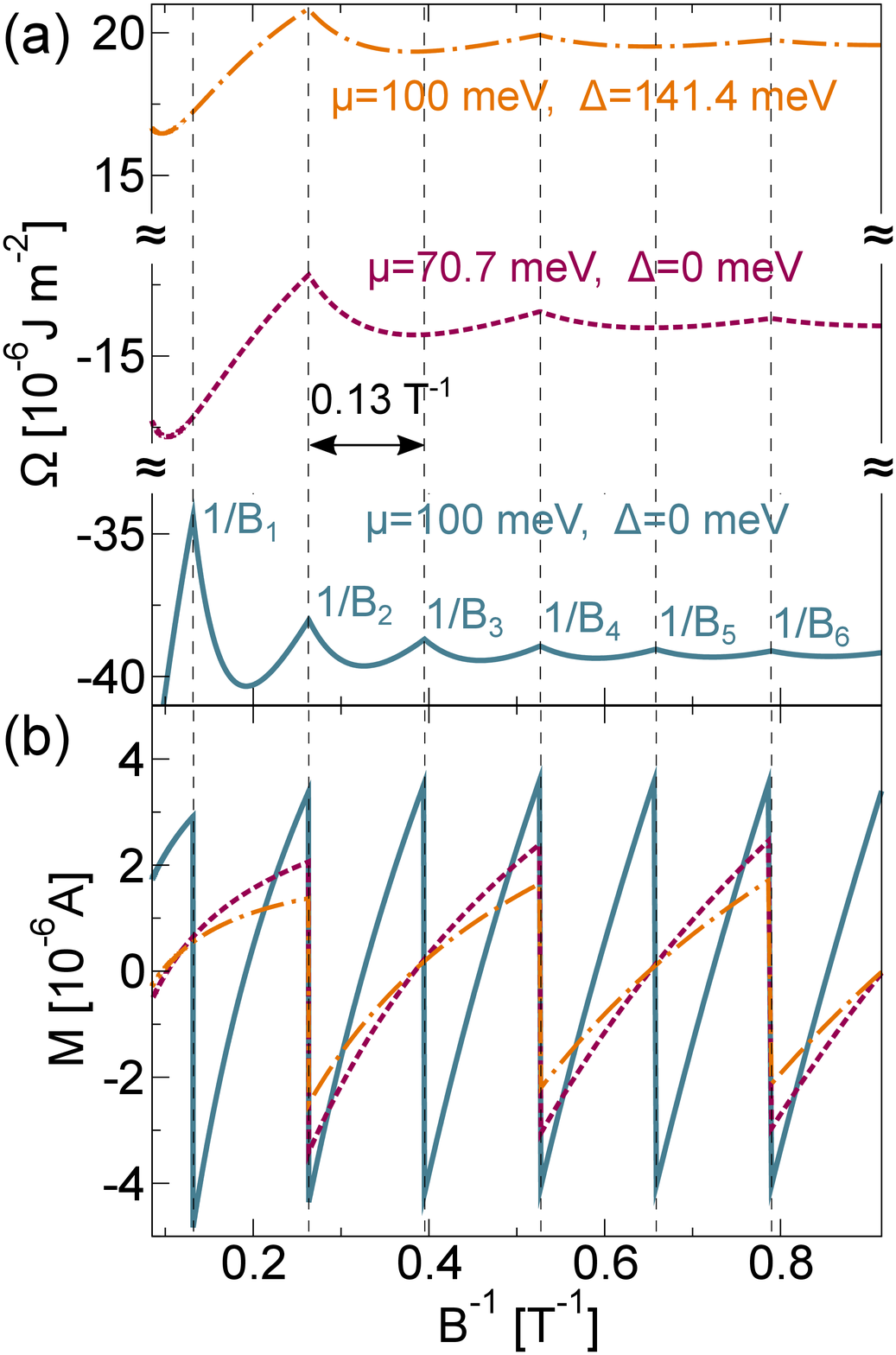}
\caption{ (Color online) The dHvA oscillations of the Dirac fermions for $\mu=100~\mathrm{meV},~\Delta=0~\mathrm{meV}$ (solid lines), $\mu=70.7~\mathrm{meV},~\Delta=0~\mathrm{meV}$ (dashed lines), and $\mu=100~\mathrm{meV},~\Delta=141.4~\mathrm{meV}$ (dash-dotted lines) in (a) thermodynamic potential and (b) magnetization. }
\label{fig:dHvADirac}
\end{center}
\end{figure}

In Fig.~\ref{fig:DiracT0}, we plot the magnetization of the massive Dirac fermions as a function of $B$ for several values of $\Delta$ at $T=0~\mathrm{K}$. We can see that the magnetization undergoes a gradual change from $M\propto-\sqrt{B}$ to $M\propto-B$ dependences with increasing $\Delta$ for $B=1-10~\mathrm{T}$, which indicates that the anomalous orbital diamagnetism for $\Delta=0$ disappears with opening the gap. In this case, the spacing of the LLs which is initially $\sqrt{|n|}$ dependence becomes constant $(\hbar\omega_c)^2/\Delta$ with increasing $\Delta$. This process can be observed by the transition from the topological to the trivial phases of undoped silicene in which the band-gap can be controlled by applying an external electric field~\cite{ezawa2012ezawa} perpendicular to the silicene plane. A similar transition is predicted in graphene with increasing temperature for the same reason, in which the $\sqrt{|n|}$ dependence of the LLs in the valence bands is responsible for the $M\propto-\sqrt{B}$ dependence. When the thermal energy becomes larger than the cyclotron energy, the effect $\sqrt{|n|}$ spacing of the LLs on the magnetization becomes no more important and thus the Dirac system shows linear response $M\propto-B$ at a high $T$.

The oscillation of magnetization or susceptibility in a uniform magnetic field, which is known as the de Haas-van Alphen (dHvA) effect, has been observed experimentally in a quasi-2D system of graphite~\cite{lukyanchuk2004phase}. Much of the theoretical studies on the magnetic oscillations in the 2D systems~\cite{champel2001dehaas,lukyanchuk2011dehaas,wright2013quantum,kishigi2014quantum} have been carried out  within the framework of a generalized Lifshits-Kosevich (LK)~\cite{lifshits1955theory} theory, which was proposed to account the magnetic oscillations in metals. In the LK theory, the dHvA effect is expressed by adding an oscillatory term to the Euler-Maclaurin formula (also known as the Poisson summation formula) for calculating the thermodynamic potential~\cite{lifshits1955theory}. 

By assuming a fixed $\mu$ and $\lambda=0$ in Eqs.~(\ref{eq:OvlowT}) and (\ref{eq:OclowT}), let us discuss the effect of band-gap on the period and amplitude of the dHvA oscillation at $T=0~\mathrm{K}$. In the bottom panel of Fig.~\ref{fig:dHvADirac}(a), we plot $\Omega$ for $\mu=100~\mathrm{meV}$ and $\Delta=0~\mathrm{meV}$ as a function of inverse magnetic field $1/B$. At several values of $1/B$ (labelled as $1/B_{\nu},~\nu=1,~2,...$), we observe peaks of $\Omega$ which indicate the local maxima of potential and the peaks are separated by a period of $0.13~\mathrm{T^{-1}}$. At $1/B_{\nu}$, the $\nu$-th LLs at the $K$ and $K^{\prime}$ valleys exactly match the chemical potential $\mu$ and thus we get 
\begin{align}
\nu=\frac{\mu^2-(\Delta/2)^2}{2\hbar{v_F}^2eB_{\nu}}=\frac{\hbar A_F}{2\pi e B_{\nu}},
\label{eq:onsager}
\end{align}
where $A_F=\pi {k_F}^2=\pi[\mu^2-(\Delta/2)^2]/(\hbar{v_F})^2$ is the area of the Fermi surface of the Dirac system. The rightmost side of Eq.~(\ref{eq:onsager}) is the relation derived by Onsager~\cite{onsager1952interpretation} to demonstrate that the dHvA oscillation can be utilized to reconstruct the Fermi surface in metals. The period of the dHvA oscillation in the massive Dirac system is given as follows~\cite{mcclure1956diamagnetism,sharapov2004magnetic}:
\begin{align}
P=\frac{1}{B_{\nu+1}}-\frac{1}{B_{\nu}}=\frac{2\hbar{v_F}^2e}{\mu^2-(\Delta/2)^2}.
\label{eq:period}
\end{align}
In the middle and upper panels of Fig.~\ref{fig:dHvADirac}(a), we plot $\Omega$ by adopting $\mu=100/\sqrt{2}~\mathrm{meV}\approx 70.7~\mathrm{meV},~\Delta=0~\mathrm{meV}$ and $\mu=100~\mathrm{meV},~\Delta=100\sqrt{2}~\mathrm{meV}\approx 141.4~\mathrm{meV}$, respectively. In the both cases, the periods of the oscillation are doubled, which is consistent with Eq.~(\ref{eq:period}). Thus, the period of the dHvA can be used to extract the value $\mu$ relative to the band gap $\Delta$. This method is originally proposed by Sharapov et al.~\cite{sharapov2004magnetic} to detect the opening of band-gap in graphene with keeping $\mu$ constant. Experimentally, the band-gap opening was observed~\cite{zhou2007substrate} in epitaxially grown graphene on SiC substrate, where $\Delta\approx0.26~\mathrm{eV}$ is observed by breaking of sublattice symmetry due to the graphene-substrate interaction. 

In Fig.~\ref{fig:dHvADirac}(b) we plot the magnetization for the corresponding values of $\mu$ and $\Delta$ provided in Fig.~\ref{fig:dHvADirac}(a), where the oscillations exhibits a sawtooth-like feature. It is known from the LK theory that the sawtooth-like oscillations is a characteristic of the dHvA effect in 2D systems~\cite{sharapov2004magnetic,champel2001dehaas,escudero2019temperature,escudero2020general}. We show that the sawtooth oscillation of $M$ can be derived from the zeta functions in Eqs. Eq.~(\ref{eq:OvlowT}) and (\ref{eq:OclowT}). By using the formula $\partial \zeta(p,q)/\partial q=-p\zeta(p+1,q)$, the magnetization is expressed analytically as follows:
\begin{align}
\begin{split}
M&=\frac{4e}{h}\Bigg[\frac{3}{2}\hbar\omega\zeta\Bigg(\frac{-1}{2},\phi\Bigg)+\mu\Bigg(\nu+\frac{1}{2}\Bigg)-\mu\widetilde{\nu}\sum_{\nu}\delta(\widetilde{\nu}-\nu)\\
&~~~-\frac{1}{2}\hbar\omega\zeta\Bigg(\frac{1}{2},\phi\Bigg)\Bigg\{\Gamma^2+\widetilde{\nu}\sum_{\nu}\delta(\widetilde{\nu}-\nu)\Bigg\}\Bigg],
\end{split}
\label{eq:sawtooth}
\end{align}
where we define $\phi\equiv\Gamma^2+\nu+1$. Thus, the sawtooth-like oscillation in the magnetization originates from the delta function at $\tilde{\nu}=\nu$ in Eq.~(\ref{eq:sawtooth}), which is the result of differentiation of the floor function in the expression of $\nu$ [$\partial\lfloor x \rfloor/\partial x=\sum_{n\in\mathbb{Z}}\delta(x-n)$]. Physically, the delta function indicates the occupations of electrons occupying discrete LLs. With the increase of temperature, impurity scattering, electron-electron interactions, and electron-phonon interactions~\cite{sharapov2004magnetic,yang2010landau,funk2015microscopic,sobol2016screening}, the LLs become broad in which the delta function is replaced by a Lorentzian function to account the broadening by the interactions. As a result, the oscillation of magnetization becomes less sharp. The effects of the broadening on the dHvA oscillation can be incorporated by the convolution of the thermodynamic potential at $T=0~\mathrm{K}$ with the distribution functions for temperature and impurities, as given in references~\cite{sharapov2004magnetic,knolle2015quantum,becker2019magnetic}.
 
We observe that in the cases of $\Delta=0$ (solid and dashed lines in Fig.~\ref{fig:dHvADirac}(b)), the smaller $\mu$ not only yields a decreasing frequency but also a weaker amplitude in the oscillation. When we consider the cases of $\sqrt{\mu^2-(\Delta/2)^2}=70.7~\mathrm{meV}$ (dashed and dash-dotted lines), the magnetization with the non-zero band-gap (dash-dotted line) produces a smaller amplitude in the oscillation than the case with zero band-gap (dashed line). The effect of $\Delta$ on the magnitude of the oscillation appears in the last term of Eq.~(\ref{eq:sawtooth}) [$\Gamma=\Delta/(2\hbar\omega)$], and therefore the opening of the band-gap decreases the amplitude of the magnetization as the functions $\zeta(-1/2,\phi)$ and  $\zeta(1/2,\phi)$ possess the same signs for a given $\phi$ (see Appendix B). The effect of SOC on the magnetic oscillation in TMDs with the Zeeman splitting will be the subject of the next work.

\subsection{Magnetization of TMDs ($\Delta\neq 0,~\lambda\neq0$)}

\begin{figure}[t]
\begin{center}
\includegraphics[width=7 cm, height=4.678 cm]{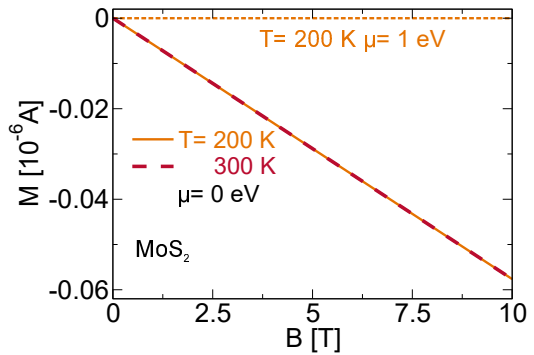}
\caption{ (Color online) Magnetization of undoped and doped ($\mu=1~\mathrm{eV}$) $\mathrm{MoS_2}$ as a function $B=0-10~\mathrm{T}$ at $T=200$ and $300~\mathrm{K}$.  }
\label{fig:MoS2Th}
\end{center}
\end{figure}

For TMDs in a magnetic field $B$ up to $\sim 10~\mathrm{T}$, the LLs that are given by Eq.~(\ref{eq:landaulls}) are approximated by $\epsilon_{n}^{\xi}\approx\xi\lambda-\Delta/2-(\hbar\omega_c)^{2}|n|/\Delta_{\xi}$ and $\epsilon_{n}^{\xi}\approx\Delta/2+(\hbar\omega_c)^{2}|n|/\Delta_{\xi}$ for $\mathrm{sgn}_{\tau}(n)=-1$ and $\mathrm{sgn}_{\tau}(n)=+1$, respectively. Thus, the LLs separation is inversely proportional to the band-gap, i.e. $(\hbar\omega_c)^{2}/\Delta_{\xi}$. This approximation is also valid for heavy Dirac fermions such as h-BN where $\Delta\approx 6~\mathrm{eV}$~\cite{kubota2007deep,kim2011synthesis,cassabois2016hexagonal} by putting $\lambda=0$. The thermodynamic potentials for TMDs in the case of $(\hbar\omega_c)^{2}/\Delta_{\xi}\ll k_B T$ are given by
\begin{align}
\begin{split}
\Omega_{-}&=-\frac{2}{\beta}\frac{eB}{h}\sum_{\xi=\pm 1}\sum_{l=1}^{\infty}\mathrm{Li}_{1-l}[-e^{\beta(\mu-\xi\lambda+\Delta/2)}]\Big(\frac{\beta}{\Delta_{\xi}}\Big)^{l}\\
&~~~\times\frac{(\hbar\omega_c)^{2l}}{l!}\frac{\mathcal{B}_{l+1}}{l+1},
\end{split}
\label{eq:TMDOvlowT}
\end{align}
and
\begin{align}
\begin{split}
\Omega_{+}&=-\frac{2}{\beta}\frac{eB}{h}\sum_{\xi=\pm 1}\sum_{l=1}^{\infty}\mathrm{Li}_{1-l}[-e^{\beta(\mu-\Delta/2)}]\Big(\frac{-\beta}{\Delta_{\xi}}\Big)^{l}\\
&~~~\times\frac{(\hbar\omega_c)^{2l}}{l!}\frac{\mathcal{B}_{l+1}}{l+1}.
\end{split}
\label{eq:TMDOclowT}
\end{align}
Note that in  Eqs.~(\ref{eq:TMDOvlowT}) and (\ref{eq:TMDOclowT}), the summations begin from $l=1$ because the terms of $l=0$ cancel with the thermodynamic potentials originated from the zeroth LLs $n=0$ (see Appendix \ref{A4} for derivation). Thus, the entropy of electrons at the zeroth LLs is not manifested in a linear $T$ dependence of as in the case of graphene. By keeping only the first term of $\Omega$, the thermodynamic potential of TMDs are given by
\begin{align}
\Omega\approx\frac{e^2{v_F}^2B^2}{6\pi}\sum_{\xi=\pm}\frac{1}{\Delta_{\xi}}\frac{\sinh\Big[\frac{\beta\Delta_{\xi}}{2}\Big]}{\cosh\Big[\frac{\beta\Delta_{\xi}}{2}\Big]+\cosh\Big[\beta\Big(\mu-\frac{\xi\lambda}{2}\Big)\Big]}.
\label{eq:TMDOlowT}
\end{align}
From Eq.~(\ref{eq:TMDOlowT}), we can see that $\Omega\propto B^2$ and therefore the magnetization is linearly proportional to $B$, which prevails only for heavy Dirac fermions.

In Fig.~\ref{fig:MoS2Th}, we plot $M$ of undoped $\mathrm{MoS_2}$ as a function of $B$ for $B=0-10~\mathrm{T}$ at $T=200~\mathrm{K}$ and $300~\mathrm{K}$, where the magnetization does not change with the increasing temperature from $T=200~\mathrm{K}$ to $300~\mathrm{K}$. Thus, even though the magnitude of magnetization in a heavy Dirac fermion decreases with the increasing band gap, the magnetization is robust for temperature. In fact, by comparing Eq.~(\ref{eq:TMDOlowT}) with (\ref{eq:Ovheavy}), we infer that the magnetizations of the heavy Dirac fermions with a given $\Delta$ at $T=0~\mathrm{K}$ and finite temperatures are equal, provided that $\Delta/2\gg k_B T$. In Fig.~\ref{fig:MoS2Th}, we also plot $M$ for a doped case ($\mu=1~\mathrm{eV}$) at $T=200~\mathrm{K}$ where the magnetization becomes zero, which demonstrates the effect of pseudospin paramagnetism~\cite{koshino2010anomalous,koshino2011singular}. By neglecting the spin-orbit coupling, the susceptibility of the massive Dirac fermion as a function of $\Delta$ and $T$ is given by 
\begin{align}
\chi=-\frac{2e^2{v_F}^2}{3\pi\Delta}\frac{\sinh\Big[\frac{\beta\Delta}{2}\Big]}{\cosh\Big[\frac{\beta\Delta}{2}\Big]+\cosh[\beta\mu]},
\label{eq:koshino}
\end{align}
which reproduce the result by Koshino and Ando ~\cite{koshino2010anomalous,koshino2011singular} with the Euler-Maclaurin formula.

\subsection{Susceptibility of grapehene and TMDs with impurity}

\begin{figure}[t]
\begin{center}
\includegraphics[width=8.2 cm, height=5.506 cm]{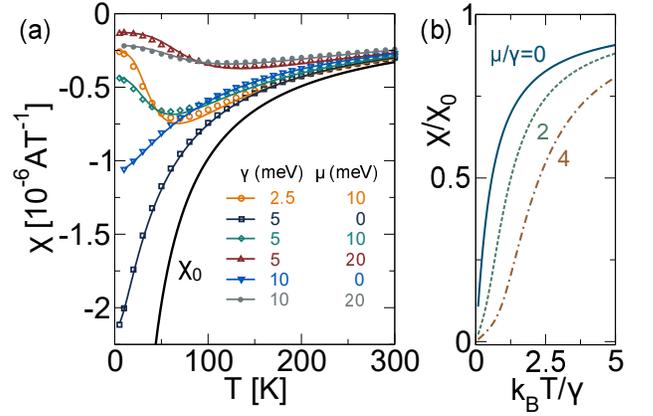}
\caption{ (Color online) The susceptibility of graphene with impurity as a function of temperature. (a) The calculation of $\chi$ as a function of $T=0-300~\mathrm{K}$ for several values of $\gamma$ and $\mu$ by the Faddeeva function (solid lines) numerical calculations (symbols). (b) The scaled susceptibility $\chi/\chi_0$ as a function of $k_B T/\gamma$ for $\mu/\gamma=0,~2,$ and $4$ with the Faddeeva function.    }
\label{fig:faddeevaT}
\end{center}
\end{figure}

\begin{figure}[t]
\begin{center}
\includegraphics[width=7 cm, height=5.381 cm]{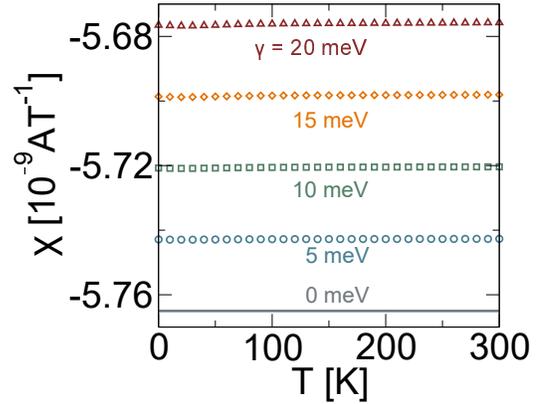}
\caption{ (Color online) The susceptibility of undoped $\mathrm{MoS_2}$ with impurity as a function of temperature $T=0-300~\mathrm{K}$ for $\gamma=0,~5,~10$, and $20~\mathrm{meV}$. }
\label{fig:MoS2I}
\end{center}
\end{figure}

Finally, we analyse the effect of impurity scattering on the orbital susceptibility of the Dirac fermions by using Eq.~(\ref{eq:impurity}) for susceptibility. In the case of graphene, we approximate the function $\mathrm{sech}^2(\beta\varepsilon/2)$ in Eq.~(\ref{eq:mcclure}) by a Gaussian function $\exp[-(C\beta\varepsilon)^2]$, where $C$ is a constant defined by $C\equiv\sqrt{\mathrm{ln}2}/[\sqrt{2}\mathrm{ln}(2+\sqrt{3})]\approx 0.447$ (see Appendix C for detail). The solution of the Voigt profile (convolution of the Gaussian with the Lorentzian functions) is given by the real part of the Faddeeva function $w(z)$ as follows \cite{olver2010nist}:
\begin{align}
V(x,y,\sigma)\equiv\frac{y}{\pi}\int_{-\infty}^{\infty}dt\frac{\exp[-t^{2}/(2\sigma^2)]}{(t-x)^2+y^2}=\mathrm{Re}[w(z)],
\label{eq:voigt}  
\end{align}
where $\sigma$ is the standard deviation of the Gaussian function, $z=(x+iy)/(\sqrt{2}\sigma)$, and $w(z)$ is the Faddeeva function defined by 
\begin{align}
w(z)\equiv e^{-z^2}\Bigg(1+\frac{2i}{\sqrt{\pi}}\int_{0}^{z}dt e^{t^2}\Bigg).
\label{eq:faddeeva}  
\end{align}
Therefore, in the presence impurity, the orbital susceptibility of graphene is approximately given by 
\begin{align}
\chi(\mu,\gamma)\approx-\frac{e^2{v_F}^2}{6\pi k_B T}\mathrm{Re}[w(z^{\prime})],
\label{eq:faddeevaxi}
\end{align}
where we define $z^{\prime}\equiv C\beta(\mu+i\gamma)$. 

In Fig. \ref{fig:faddeevaT}(a), we plot the susceptibility graphene as a function of temperature for several values of $\gamma$ and $\mu$ by using Eq.~(\ref{eq:faddeevaxi}) (lines) as well as by numerical calculation of the convolution by using the $\mathrm{sech}^2(\beta\varepsilon/2)$ function (symbols). We can see that the approximation with the Fadddeeva function is in a good agreement with the numerical calculation. For comparison, we show the susceptibility of undoped graphene without impurity $\chi_0$ by putting $\mu=0$ in Eq.~(\ref{eq:mcclure}) [$\chi_0=-(e v_F)^2/(6\pi k_B T)$], which is inversely proportional to the temperature. From Fig. \ref{fig:faddeevaT}(a), the susceptibility for non-zero $\gamma$ is finite as $T\rightarrow 0~\mathrm{K}$, which shows that the anomalous diamagnetism in graphene disappears by introducing the impurity effect. In the cases of $\mu\neq 0$, we observe minimum values of $\chi$ at finite temperatures. For $\gamma=5~\mathrm{meV}$, the minimum value becomes smaller and shift to the higher temperature as we increase $\mu$ from $10~\mathrm{meV}$ to $20~\mathrm{meV}$. The present method reproduces the calculation by Nakamura and Hirasawa~\cite{nakamura2008electric}, where the susceptibility of graphene with impurity is approximated by the Sommerfeld expansion and also shows the minimum values in the susceptibility as a function of temperature. In Fig.~\ref{fig:faddeevaT}(b), we plot $\chi/\chi_0$ as a function of $k_B T/\gamma$. For a given ratio $\mu/\gamma$, the curves shown in Fig.~\ref{fig:faddeevaT}(a) follow the scaling law shown in Fig.~\ref{fig:faddeevaT}(b). Therefore, the advantage of using the Faddeeva function is that the susceptibility of graphene in the presence of the impurity scattering is approximately scaled by the function $\mathrm{Re}[w(z^{\prime})]$.

In Fig.~\ref{fig:MoS2I}, we numerically calculate the susceptibility of undoped $\mathrm{MoS_2}$ as a function of $T$ for several values of $\gamma$. Here, $\chi$ does not change with increasing $T$. As we increase $\gamma$, the magnitude of $\chi$ decreases with the same rate, which means that for a given temperature, the susceptibility decreases linearly as a function of $\gamma$.

\section{Conclusion}

In this study, the analytical expressions of thermodynamic potentials and magnetizations of the Dirac fermions are derived by using the technique of zeta function regularization. There are four main results obtained in the study: (1) the analytical formula reproduce the Langevin fitting for the magnetization of graphene for two limits of $\hbar\omega_c\gg k_BT$ and $\hbar\omega_c\ll k_BT$. (2) We derive the formula for the magnetization of heavy Dirac fermions and show that the magnetization is robust with respect to temperature and impurity scattering. (3) The scaling law for the susceptibility of graphene with impurity scattering can be approximated by the real part of the Faddeeva function, and (4) the gap effect on the dHvA oscillation at $T=0~\mathrm{K}$ are discussed from the property of the zeta function in the thermodynamic potential. All results by taking zeta function regularization reproduces the previous work by taking some limits. Thus, the zeta function regularization is justiied without any exceptions.

\section{Acknowledgement}

FRP acknowledges MEXT scholarship. MSU acknowledges JSPS KAKENHI Grant Number JP18J10199. RS acknowledges JSPS KAKENHI Grant Number JP18H01810.


\begin{appendix}

\renewcommand\thefigure{\thesection.\arabic{figure}} 

\section{Derivation of Eqs.~(8), (13), (16), and (29)}

\subsection{Derivation of Eq.~(8)}\label{A1}

By substituting $\epsilon_{n}^{\xi}=\xi\lambda/2-\hbar\omega_c\sqrt{|n|+{\Gamma_\xi}^2}$ for $n\leqslant 0$ in the expression of $\Omega_{-}$, we get
\setcounter{equation}{0}
\begin{align}
\Omega_{-}=\frac{eB}{h}\sum_{\xi=\pm }\Bigg[\sum_{n=0}^{\infty}+\sum_{n=1}^{\infty}\Bigg]\Bigg[\frac{\xi\lambda}{2}-\hbar\omega_c\sqrt{n+{\Gamma_\xi}^2}-\mu\Bigg],
\label{eq:8a}
\end{align}
where the summation operator which begins from $n=0$ ($n=1$) indicates the sum of the LLs at the $K$ ($K^{\prime}$) valley. Now, let us shift the index of the summation from $n=1$ to $n=0$ for the term $-\hbar\omega_c\sqrt{n+{\Gamma_{\xi}}^2}$ as follows:
\begin{align}
\begin{split}
\Omega_{-}^{(e)}&=\frac{eB}{h}\sum_{\xi=\pm }\Bigg[\frac{\xi\lambda}{2}+2\sum_{n=1}^{\infty}\frac{\xi\lambda}{2}-\mu-2\sum_{n=1}^{\infty}\mu\\
&~~~+\frac{\Delta_\xi}{2}-2\hbar\omega_c\sum_{n=0}^{\infty}\sqrt{n+{\Gamma_\xi}^2}\Bigg].
\end{split}
\label{eq:8b}
\end{align}
The first and second summations are expressed by the Riemann zeta function $\zeta(p)=\sum_{k=1}^{\infty}{1}/{k^p}$, while the third summation is expressed by the Hurwitz zeta function [Eq.~(\ref{eq:zeta})]. By using $\sum_{n=1}^{\infty}=\zeta(0)=-1/2$ for the first and second summations, we derive the $\Omega_{-}$ as given by Eq.~(\ref{eq:OvlowT}).

\subsection{Derivation of Eq.~(13)}\label{A2}

By substituting $\epsilon_{n}^{\xi}=\xi\lambda/2+\hbar\omega_c\sqrt{|n|+{\Gamma_\xi}^2}$ for $n\geqslant 0$, $\Omega_{+}^{(e)}$ is given by
\begin{align}
\begin{split}
\Omega_{+}^{(e)}&=\frac{eB}{h}\sum_{\xi=\pm }\Bigg[\sum_{n=1}^{\nu_{\xi}}+\sum_{n=0}^{\nu_{\xi}}\Bigg]\Bigg[\frac{\xi\lambda}{2}+\hbar\omega_c\sqrt{n+{\Gamma_\xi}^2}-\mu\Bigg]\\
&~~~\times\Theta(\mu-\Delta/2)\\
&=\frac{eB}{h}\sum_{\xi=\pm }\Bigg[ \Bigg(\frac{\xi\lambda}{2}-\mu\Bigg)(2\nu_{\xi}+1) \\
&~~~-\frac{\Delta_{\xi}}{2}+2\hbar\omega_c\sum_{n=0}^{\nu_{\xi}}\sqrt{n+{\Gamma_{\xi}}^2}\Bigg]\Theta(\mu-\Delta/2).
\end{split}
\label{eq:13}
\end{align}
Here, the summation operator which begins from $n=1$ ($n=0$) indicates the summations of the LLs at the $K$ ($K^{\prime}$) valley. By expressing the summation of $n$ as a difference of two zeta functions [see Eq.~(\ref{eq:difference})], we get Eq.~(\ref{eq:OclowT}) in the main text. With the same method, by using  $\epsilon_{n}^{\xi}=\xi\lambda/2-\hbar\omega_c\sqrt{|n|+{\Gamma_\xi}^2}$ for $n\leqslant 0$, we can derive Eq.~(\ref{eq:OhlowT}) for $\Omega_{-}^{(h)}$.

\subsection{Derivation of Eq.~(16)}\label{A3}

The thermodynamic potential $\Omega_{+}^{\prime}$ for $T>0~\mathrm{K}$ is calculated by convolution of $\Omega_{+0}^{\prime}\equiv \Omega_{+}^{\prime}(T=0~\mathrm{K})$ with $(-\partial f/\partial \varepsilon)=\beta\mathrm{sech}^2[\beta(\varepsilon-\mu)/2]/4$, where $f(\varepsilon)$ is the Fermi distribution function~\cite{mcclure1956diamagnetism,sharapov2004magnetic}. By using $\epsilon_{n}^{\xi}=\hbar\omega_c\sqrt{|n|+\Gamma^2}$, $\Omega_{+0}^{\prime}$ is given by
\begin{align}
\begin{split}
\Omega_{+0}^{\prime}&=2g_s\frac{eB}{h}\sum_{n=\nu}^{\infty}[\epsilon_{n}^{\xi}-\mu]\\
&=2g_s\frac{eB}{h}\Bigg[\hbar\omega_c\zeta\Big(-\frac{1}{2},\Gamma^2+\nu+1\Big)+\Big(\nu-\frac{1}{2}\Big)\mu\Bigg].
\end{split}
\label{eq:16a}
\end{align}
Here, in the second line of Eq.~(\ref{eq:16a}) we use $\sum_{n=\nu}^{\infty}=\zeta(0)-\sum_{n=1}^{\nu-1}=-\nu+1/2$. Since $\Gamma^2+\nu\approx\mu^2/(\hbar\omega_c)^2$ and by considering $\mu\gg\hbar\omega_c$, the zeta function $\zeta(-1/2,\Gamma^2+\nu+1)$ can be approximated by using Eq.~(\ref{eq:asymptotic}). $\Omega_{+0}^{\prime}$ and $\Omega_{+}^{\prime}$ are given by
\begin{align}
\Omega_{+0}^{\prime}(\mu)\approx\frac{1}{2\pi}\frac{g_s}{(\hbar v_F)^2}\Bigg[ \frac{\mu^3}{3}-\frac{\Delta^2\mu}{4}-\mu(\hbar\omega_c)^2-\frac{(\hbar\omega_c)^4}{24\mu}\Bigg].
\label{eq:16b}
\end{align}
and
\begin{align}
\Omega_{+}^{\prime}(\mu)=\frac{\beta}{4}\int_{-\infty}^{\infty}~d\varepsilon~\Omega_{+0}^{\prime}(\varepsilon)~\mathrm{sech}^{2}\Big[\frac{\beta}{2}(\varepsilon-\mu)\Big],
\label{eq:16c}
\end{align}
respectively, where $\Omega_{+0}^{\prime}(\varepsilon)$ is given by substituting $\mu$ in Eq.~(\ref{eq:16b}) to variable $\varepsilon$. $\Omega_{+0}^{\prime}(\varepsilon)$ is an odd of $\varepsilon$. In the case of $\mu\ll k_B T$, the function $\mathrm{sech}^2[\beta(\varepsilon-\mu)/2]$ can be approximated as an even function. Therefore, we get $\Omega_{+}^{\prime}(\mu)\approx 0$ for $T>0~\mathrm{K}$ and $\mu\ll k_B T$.

Now, by expanding the logarithmic and exponential functions in the expression of thermodynamic potential for $\hbar\omega_c\sim\Delta\ll k_B T$, and by using $\epsilon_{-n}^{\xi}=-\hbar\omega_c\sqrt{|n|+\Gamma^2}$, $\Omega_{-}$ is given by
\begin{align}
\begin{split}
\Omega_{-}&=\frac{-g_s}{\beta}\frac{eB}{h}\Bigg[\sum_{n=0}^{\infty}+\sum_{n=1}^{\infty}\Bigg]\sum_{k=1}^{\infty}\frac{(-1)^{k-1}}{k}e^{\beta \mu k}\exp(-\beta\epsilon_{-n}^{\xi} k)\\
&=\frac{g_s}{\beta}\frac{eB}{h}\Bigg[\sum_{n=0}^{\infty}+\sum_{n=1}^{\infty}\Bigg]\sum_{k=1}^{\infty}\frac{(-e^{\beta\mu})^{k}}{k}\sum_{l=0}^{\infty}\frac{(-\beta k \epsilon_{-n}^{\xi} )^l}{l!}\\
&=\frac{g_s}{\beta}\frac{eB}{h}\sum_{l=0}^{\infty}\sum_{k=1}^{\infty}\frac{(-e^{\beta \mu })^k}{k^{1-l}}\frac{(\beta \hbar \omega_c)^l }{l!}\\
&~~~\times\Bigg[2\sum_{n=0}^{\infty}(n+\Gamma^{2})^{l/2}-\Gamma^{l}\Bigg]\\
&=\frac{2 g_s eB}{\beta h}\sum_{l=0}^{\infty}\mathrm{Li}_{1-l}(-e^{\beta\mu})\frac{(\beta \hbar \omega_c)^l }{l!}\Bigg[\zeta\Big(\frac{-l}{2},\Gamma^2\Big)-\frac{\Gamma^l}{2}\Bigg].
\end{split}
\label{eq:16d}
\end{align} 
In the third line of Eq.~(\ref{eq:16d}), we switch the order of summations with indices $k$ and $l$ in order to express the $\mu$ dependence of $\Omega_{-}$ in term of the polylogarithm function $\sum_{k=1}^{\infty}(-e^{\beta\mu})^{k}/k^{1-l}=\mathrm{Li}_{1-l}(-e^{\beta\mu})$, and we shift the sum operator which begins from $n=1$ to $n=0$ in order to adjust the Hurwitz zeta function. Similarly, the expression for $\Omega_{+}$ is given by
\begin{align}
\Omega_{+}=\frac{2 g_s eB}{\beta h}\sum_{l=0}^{\infty}\mathrm{Li}_{1-l}(-e^{\beta\mu})\frac{(-\beta \hbar \omega_c)^l }{l!}\Bigg[\zeta\Big(\frac{-l}{2},\Gamma^2\Big)-\frac{\Gamma^l}{2}\Bigg].
\label{eq:16e}
\end{align}
When we add Eqs.~(\ref{eq:16d}) and (\ref{eq:16e}), the terms odd $l$ disappear, while the terms for $l=0,~2,~4,...$ are doubled. By expressing $l=2\ell$, we get Eq.~(\ref{eq:OhighT2}).

\subsection{Derivation of Eq.~(29)}\label{A4}

The $\Omega_{-}$ for TMDs is given as follows:
\begin{align}
\Omega_{-}=-\frac{1}{\beta}\frac{eB}{h}\sum_{\xi=\pm}\Bigg[\sum_{n=0}^{\infty}+\sum_{n=1}^{\infty}\Bigg]\mathrm{ln}[1+\exp\{-\beta(\epsilon_{-n}^{\xi}-\mu)\}].
\label{eq:29a}
\end{align}
By using $\epsilon_{-n}^{\xi}=\xi\lambda-\Delta/2-(\hbar\omega_c)^{2}|n|/\Delta_{\xi}$ and separating the term $n=0$ from $n=1$ to $\infty$ in the sum of $n=0$ to $\infty$, we have
\begin{align}
\begin{split}
\Omega_{-}&=-\frac{1}{\beta}\frac{eB}{h}\sum_{\xi=\pm}\mathrm{ln}[1+e^{\beta(\mu-\xi\lambda+\Delta/2)}]\\
&~~~-\frac{2}{\beta}\frac{eB}{h}\sum_{\xi=\pm}\sum_{n=1}^{\infty}\mathrm{ln}[1+e^{\beta\{\mu-\xi\lambda+\Delta/2+(\hbar\omega_c)^{2}n/\Delta_{\xi}\}}]\\
&\equiv \Omega_{-}^{\prime}+\Omega_{-}^{\prime\prime}.
\end{split}
\label{eq:29b}
\end{align}
Here, we define $\Omega_{-}^{\prime}$ and $\Omega_{-}^{\prime\prime}$ as the thermodynamic potentials for the zeroth LL at the $K$ valley and $n\leqslant -1$ LLs, respectively. By expanding the logarithmic and exponential functions in the expression of $\Omega_{-}^{\prime\prime}$, we get
\begin{align}
\begin{split}
\Omega_{-}^{\prime\prime}&=\frac{2}{\beta}\frac{eB}{h}\sum_{\xi=\pm}\sum_{k=1}^{\infty}\frac{[-e^{(\mu-\xi\lambda+\Delta/2)}]^k}{k}\\
&~~~\times\sum_{l=0}^{\infty}\Big(\frac{\beta k }{\Delta_{\xi}}\Big)^{l}\frac{(\hbar\omega_c)^{2l}}{l!}\sum_{n=1}^{\infty}n^l\\
&=\frac{2}{\beta}\frac{eB}{h}\sum_{\xi=\pm}\sum_{l=0}^{\infty}\mathrm{Li}_{1-l}[-e^{\beta(\mu-\xi\lambda+\Delta/2)}]\Big(\frac{\beta}{\Delta_{\xi}}\Big)^{l}\\
&~~~\times\zeta(-l)\\
&=\sum_{l=0}^{\infty}\Omega_{l}^{\prime\prime}
\end{split}
\label{eq:29c}
\end{align}
Because $\mathrm{Li}_{1}(z)=-\mathrm{ln}(1-z)$ and $\zeta(0)=-1/2$, we get $\Omega_{0}^{\prime\prime}=eB/(\beta h)\sum_{\xi=\pm 1}\mathrm{ln}[1+e^{(\mu-\xi\lambda+\Delta/2)}]=-\Omega_{-}^{\prime}$. As a result, only the terms for $l\geqslant 1$ survives in the final expression of $\Omega_{-}$ [Eq.~(\ref{eq:TMDOvlowT})]. Similarly, by using $\epsilon_{n}^{\xi}=\Delta/2+(\hbar\omega_c)^{2}|n|/\Delta_{\xi}$ for $n\geqslant 0$, we can derive Eq.~~(\ref{eq:TMDOclowT}).

\section{Numerical calculations of the zeta function}

\begin{figure}[t]
\setcounter{figure}{0}
\begin{center}
\includegraphics[width=8 cm, height=6.149 cm]{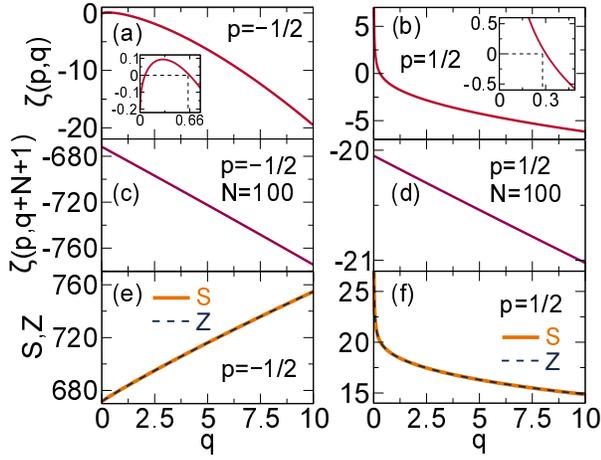}
\caption{ (Color online) Plot of $\zeta(p,q)$ with $q=1-10$ for (a) $p=-1/2$, (b) $p=1/2$. In the insets of (a) and (b) we show that the values of $\zeta(p,q)$ for $p=-1/2$ and $p=1/2$ are negative for $q>0.66$ and $q>0.3$, respectively. Plot of $\zeta(p,q,N+1)$ with $N=100$ for (c) $p=-1/2$, (d) $p=1/2$. The comparison between the functions $S(p,q,N)$ and $Z(p,q,N)$ for given $p$, $q$, and $N$ are shown in (e) and (d).    }
\label{fig:Hurwitzf}
\end{center}
\end{figure}

\begin{figure}[t]
\begin{center}
\includegraphics[width=8 cm, height=4.587 cm]{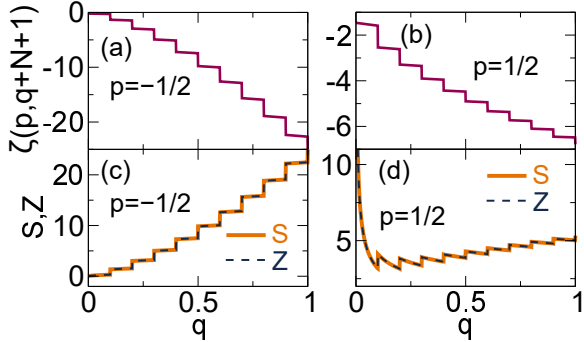}
\caption{ (Color online) Plot of $\zeta(p,q+N+1)$ with $N=10\lfloor q\rfloor$ for (a) $p=-1/2$, (b) $p=1/2$. The comparison between the functions $S(p,q,N)$ and $Z(p,q,N)$ for given $p$, $q$, and $N$ are shown in (e) and (f). }
\label{fig:Hurwitzf2}
\end{center}
\end{figure}

In Fig.~(\ref{fig:Hurwitzf}) (a) and (b), we plot the zeta function $\zeta(p,q)$ for $p=1/2$ and $p=-1/2$, respectively. The value of $\zeta(1/2,q)$ is negative and deceases monotonically for $q>0.66$ as shown in the inset in (a). The value of $\zeta(-1/2,q)$ diverges at $q=0$ and change signs at $q\approx 0.3$. In (c) and (d), we substitute $q$ to $q+N+1$ and take $N=100$ for explaining the change of $\zeta(p,q)-\zeta(p,q+N+1)$. In (e) and (f) we compare the functions $S(p,q,N)\equiv\sum_{n=0}^{N}(n+q)^{-p}$ and $Z(p,q,N)\equiv\zeta(p,q)-\zeta(p,q,N+1)$, similar to the left-hand and right-hand sides of Eq.~(\ref{eq:difference}), respectively. It is observed that the two functions exactly identical for $q=0$ to $10$. It is noted that both the functions $S$ and $Z$ are continuous and do not explain the oscillatory behaviour of the thermodynamic potential in the dHvA effect. By changing the constant $N$ to $10\lfloor q \rfloor$ for an example, the function $\zeta(p,q,N+1)$ shows step-like behaviour as shown in Fig.~\ref{fig:Hurwitzf2}(a) and (b) because of the nature of the function $\lfloor q \rfloor$. In (c) and (d), we compare the functions $S(p,q,N)$ and $Z(p,q,N)$ for $p=1/2$ and $p=-1/2$, respectively. As in the previous case, the two functions match each other. Therefore, the analytical expressions of the thermodynamic potentials for doped a Dirac fermion is numerically verified.

\section{Approximation of $\mathrm{sech}^2(\beta\epsilon/2)$ to a Gaussian function }

\begin{figure}[h]
\setcounter{figure}{0}
\begin{center}
\includegraphics[width=7 cm, height=5.043 cm]{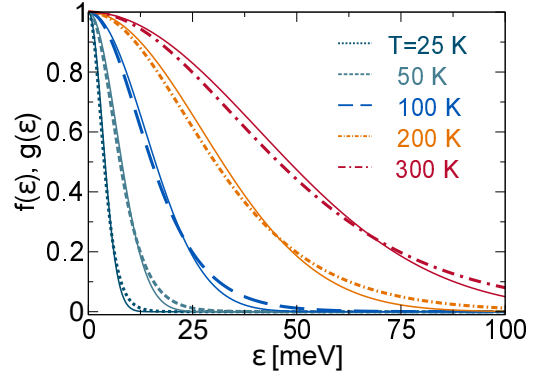}
\caption{ (Color online) Comparison between the functions $f(\varepsilon)$ and $g(\varepsilon)$ (thin solid lines) for $T=25,~50,~100,~200$ and $300~\mathrm{K}$. }
\label{fig:Gaussian}
\end{center}
\end{figure}

For given secant-hyperbolic and the Gaussian distributions, $F(\varepsilon)\equiv\mathrm{sech}(\varepsilon/W)$ and $G(\varepsilon)\equiv \exp[-\varepsilon^2/(2\sigma^2)]$, respectively, the half-width of the distributions are given by $\mathrm{HW}_{F}=\mathrm{ln}(2+\sqrt{3})W$ and $\mathrm{HW}_{G}=\sqrt{2\mathrm{ln}(2)}\sigma$. By solving $\mathrm{HW}_{F}=\mathrm{HW}_{G}$ and choosing $W=2/\beta$, the Gaussian approximation for the function $f(\varepsilon)\equiv \mathrm{sech}^2(\beta\varepsilon/2)$ is given by $g(\varepsilon)\equiv \exp[-(C\beta\varepsilon)^2]$, where $C=\sqrt{\mathrm{ln}2}/[\sqrt{2}\mathrm{ln}(2+\sqrt{3})]\approx 0.447$ as defined in the main text. In Fig.~\ref{fig:Gaussian}, we compare $f(\varepsilon)$ and $g(\varepsilon)$ (thin solid lines) for several values of temperature. The distribution $g(\varepsilon)$ has a smaller tail compared with $f(\varepsilon)$, which is the origin of discrepancies between the numerical calculation and the Faddeeva approximation in the calculation of $\chi(\mu,\gamma)$.

\end{appendix}

\end{document}